\documentclass[twocolumn,showpacs,prb,amsmath,amstex,amssymb,citeautoscript,longbibliography]{revtex4-1}
\pdfoutput=1
\usepackage{natbib}
\usepackage[english]{babel}
\usepackage{letltxmacro}
\usepackage{latexsym}
\LetLtxMacro{\ORIGselectlanguage}{\selectlanguage}
\makeatletter
\DeclareRobustCommand{\selectlanguage}[1]{%
  \@ifundefined{alias@\string#1}
    {\ORIGselectlanguage{#1}}
    {\begingroup\edef\x{\endgroup
       \noexpand\ORIGselectlanguage{\@nameuse{alias@#1}}}\x}%
}
\newcommand{\definelanguagealias}[2]{%
  \@namedef{alias@#1}{#2}%
}
\makeatother
\definelanguagealias{en}{english}
\definelanguagealias{English}{english}
\usepackage{graphicx}
\usepackage{amsmath}
\usepackage{amsfonts}
\usepackage{amssymb}
\usepackage{bm}
\usepackage{color}
\usepackage{soul} 
\usepackage{amssymb}
\usepackage{wasysym}
\usepackage{dsfont}
\usepackage{float}

\usepackage{hyperref}
\hypersetup{
    bookmarks=false,         
    unicode=false,          
    pdftoolbar=false,        
    pdfmenubar=true,        
    pdffitwindow=false,     
    pdfstartview={FitH},    
    pdftitle={Loschmidt Echo in Many-Body Localized Phase},    
    pdfauthor={M. Serbyn, D. Abanin},     
    pdfsubject={},   
    pdfcreator={M. Serbyn},   
    pdfproducer={}, 
    pdfkeywords={many-body localization} {orthogonality catastrophe} {disordered systems}, 
    pdfnewwindow=true,      
    colorlinks=true,       
    linkcolor=black,          
    citecolor=blue,        
    filecolor=magenta,      
    urlcolor=blue           
}

\newcommand{\be}{\begin{equation}}
\newcommand{\ee}{\end{equation}}
\newcommand{\bea}{\begin{eqnarray}}
\newcommand{\eea}{\end{eqnarray}}

\newcommand{\corr}[1]{\langle{ #1}\rangle}

\begin{document}
\title{Loschmidt Echo in Many-Body Localized Phase}
\author{Maksym Serbyn$^{1,3}$, Dmitry A. Abanin$^{2,3}$}
\affiliation{$^1$ Department of Physics, University of California, Berkeley, California 94720, USA}
\affiliation{$^2$ Department of Theoretical Physics, University of Geneva, 24 quai Ernest-Ansermet, 1211 Geneva, Switzerland}
\affiliation{$^3$ Kavli Institute for Theoretical Physics, University of California, Santa Barbara, CA 93106, USA}
\date{\today}
\begin{abstract}
The Loschmidt echo, defined as the overlap between quantum wave function evolved with different Hamiltonians, quantifies the sensitivity of quantum dynamics to perturbations and is often used as a probe of quantum chaos. In this work we consider the behavior of the Loschmidt echo in the many body localized phase, which is characterized by emergent local integrals of motion, and provides a generic example of non-ergodic dynamics. We demonstrate that the fluctuations of the Loschmidt echo decay as a power law in time in the many-body localized phase, in contrast to the exponential decay in few-body ergodic systems. We  consider the spin-echo generalization of the Loschmidt echo, and argue that the corresponding correlation function saturates to a finite value in localized systems. Slow, power-law decay of  fluctuations of such spin-echo-type overlap is related to the operator spreading and is present only in the many-body localized phase, but not in a non-interacting Anderson insulator. While most of the previously considered probes of dephasing dynamics could be understood by approximating physical spin operators with local integrals of motion, the Loschmidt echo and its generalizations crucially depend on the full expansion of the physical operators via local integrals of motion operators, as well as operators which flip local integrals of motion. Hence, these probes allow to get insights into the relation between physical operators and local integrals of motion, and access the operator spreading in the many-body localized phase.
\end{abstract}
\maketitle

\section{Introduction}

Despite intense theoretical studies, there remain many open questions about thermalization and emergence of statistical mechanics in quantum many-body systems. In classical many-body system  thermalization is intimately related to the chaotic behavior. Chaos in classical systems originates from the non-linearity of the classical equations of motion. Such non-linearity generally leads to a divergence of two trajectories which were initially close to each other in the phase space. The Lyapunov exponent, which sets the inverse timescale for the divergence of trajectories, is a convenient measure of classical chaotic behavior. 

In quantum systems, relation between thermalization and chaotic behavior is much less clear. The ``quantum chaos'' in a few-body quantum systems  is usually probed by the level statistics. While being a powerful probe, the level statistics provides a ``yes/no'' answer, being Wigner-Dyson (Poisson) in the ergodic (integrable) phase. At the same time, level statistics gives little insights into timescales on which thermalization emerges. Furthermore, the naive generalization of the Lyapunov exponent to the quantum systems fails. Indeed, the quantum dynamics is generated by a linear unitary operator $U=e^{-i Ht}$, and hence the overlap between different wave functions evolved with the same unitary operator remains constant in time. 

The Loschmidt echo offers an alternative way to define an analogue of Lyapunov exponent in  quantum systems. In the  Loschmidt echo setup one measures the overlap of the same wave function that was evolved with \emph{different} Hamiltonians.  More specifically, starting from an  initial state $|\psi_0\rangle$, one considers the decay of the overlap function
\begin{equation}\label{Eq:S-def}
  S(t)
  =
  \corr{\psi_0|e^{i(H_0+V)t} e^{-iH_0t}|\psi_0},
\end{equation}
 where $H_0$ is the unperturbed Hamiltonian, and $V$ is usually a local perturbation. The Loschmidt echo has been studied extensively both in the context of a few body~\cite{Beenakker02,Prosen03,Goussev09} and many-body systems~\cite{Pastawski04,Santos14,Ho15}, in particular see reviews~\cite{Prosen06,LEcho} and references therein. In  ergodic systems Loschmidt echo is believed to decay exponentially $|S(t)|^2\sim e^{-\Gamma t}$, where $\Gamma$ can be directly related to the Lyapunov exponent of the classical system within the semiclassical approach.

In this work we consider the behavior of the Loschmidt echo in many-body localized (MBL) systems. MBL phase provides a generic mechanism to avoid thermalization and break ergodicity~\cite{Basko06,Mirlin05,OganesyanHuse,PalHuse}.  The MBL phase can be characterized by the emergence of the extensive number of local integrals of motion~\cite{Serbyn13-1,Huse13}. These local integrals of motion (LIOM) do not relax, and dynamics is limited to the accumulation of random phases of eigenstates with different configuration of LIOMs, usually referred to as ``dephasing''. Dephasing dynamics in the MBL phase leads to the logarithmic spreading of entanglement~\cite{Moore12,Serbyn13-2,Vosk13} and a power-law relaxation of local observables~\cite{Serbyn14,Serbyn_14_Deer}. 

There exists an increasing number of experimental  realizations of MBL phase in systems of cold atoms~\cite{DeMarco15,Bloch15,Bloch16,Bloch16-2} and in long-range interacting ion chains~\cite{Monroe16}. However, most of the evidence for the MBL phase consists of the absence of complete relaxation in the presence of interactions, and characteristic signatures of the MBL dynamics were not yet observed~(see however recent experiments~\cite{Monroe16,Capp16}). 
While measuring  entanglement spreading experimentally is generally a very hard problem, 
the same dephasing dynamics could be detected in the relaxation of observables in a global quench~\cite{Serbyn14}, modified spin-echo type setups,~\cite{Serbyn_14_Deer} quantum revivals,~\cite{Vasseur14} and other dynamical experimental signatures of the MBL phase.~\cite{Bahri,Nandkishore14, Johri14, Vasseur14}

 In this paper we propose \emph{fluctuations} of Loschmidt echo as an alternative probe of dephasing dynamics, and demonstrate that they decay as a power-law in the MBL phase, saturating at the value that is exponentially small in the system size. At the same time, we show that the decay of overlap $S(t)$ itself, contrary to the claims of Ref.~\onlinecite{Huber16}  does not probe the dephasing dynamics of the MBL phase, but instead gives information about statistics of single particle energies. We also note that the Loschmidt echo was also studied in Ref.~\onlinecite{Torres-15} for the case when operator $V$ is a global perturbation in the MBL and ergodic phases.
 
 There are important differences between fluctuations of Loschmidt echo and other proposed probes. In contrast to the majority of other probes, the Loschmidt echo is sensitive to presence of multiple terms in the expansion of a \emph{local operator} over local integrals of motion. Hence, Loschmidt echo and its modifications can provide direct insights into the structure of local integrals of motion. In what follows we show that Loschmidt echo predominantly probes the diagonal part of the operator. In addition, we also study a spin-echo type modification of the Loschmidt echo protocol. We show that it exhibits qualitatively different behavior, saturating to a finite value in the localized phase. The fluctuations of spin echo probe the operator spreading of the off-diagonal part of local operators in the many-body localized phase. 

The paper is organized as follows: in the next section we describe the general setup for the measurement  of the  Loschmidt echo and explain its relation to the orthogonality catastrophe. We also introduce an XXZ spin chain as a specific model of the many-body localized phase and briefly review its description in terms of local integrals of motion. Next, in Section~\ref{Sec:overlap} we consider the behavior of the overlap function analytically and numerically.  Section~\ref{Sec:echo} relates the overlap function in the  spin-echo protocol to the operator spreading. Finally, in Section~\ref{Sec:summary} we summarize our results, and discuss similarities and differences between orthogonality catastrophe and other probes of dynamics in the MBL phase. Appendices provide more details on the averaged overlap $S(t)$ and behavior of spin-echo fluctuations.

\section{General setup and microscopic model}
\begin{figure}[bt]
\begin{center}
\includegraphics[width=0.95\columnwidth]{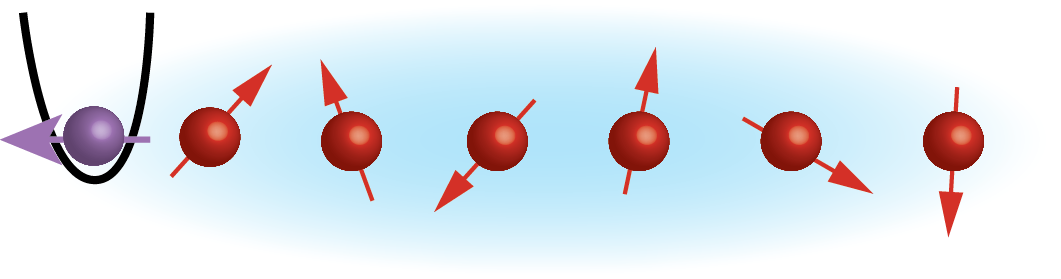}
\caption{ \label{Fig:setup} Cartoon of the setup implementing orthogonality catastrophe in the cold atom setting. The spin-1/2 impurity on the left is coupled to the spin-chain via the Zeeman interaction and has no internal dynamics. If impurity spin is initialized along the $x$-direction, the expectation value of $\sigma_\text{imp}^x$ at time $t$ gives the real part of the overlap function. 
}
\end{center}
\end{figure}

Naively the overlap function defined in Eq.~(\ref{Eq:S-def}) involves evolution of the initial wave function with two Hamiltonians, $H_0$ and $-H_0-V$. However, it can be naturally accessed  via real-time dynamics of orthogonality catastrophe setup, as proposed in Ref.~\onlinecite{Knap12}. 
In particular, let us consider an impurity coupled to a system, which is chosen to be a spin chain, as sketched in Fig.~\ref{Fig:setup}. We assume that the impurity spin has no internal dynamics, and is interacting only with its neighboring spin via Zeeman-type interaction, 
\begin{equation}\label{Eq:Coupling}
 H_c = \frac12 (1+\sigma^z_\text{imp})V,
\end{equation}
where  $V$ is an operator acting on the system's degrees of freedom, and  $ \sigma^\alpha_\text{imp}$ denotes a corresponding Pauli operator acting on the impurity. 

Under the assumption that the impurity spin has no internal dynamics it is possible to extract the overlap function~(\ref{Eq:S-def}) from a local measurement of the impurity spin. Let us prepare the full system initially in the product state,
\begin{equation}\label{Eq:psi-in}
 |\Psi \rangle =|\rightarrow\rangle_\text{imp} \otimes |\psi_0\rangle
\end{equation} 
where $|\rightarrow\rangle_\text{imp}$ denotes the state with impurity spin along $x$-axis. Evolving the state $ |\Psi \rangle$ with the full Hamiltonian $H_\text{f}=H_0+ H_c$, we obtain:
\begin{multline}\label{Eq:psi-in-t}
 |\Psi(t) \rangle  = e^{-iH_\text{f} t} |\Psi \rangle =\frac{1}{\sqrt2}|\uparrow\rangle_\text{imp} \otimes  e^{-i(H_0+V) t} |\psi_0\rangle
\\
+\frac{1}{\sqrt2}|\downarrow\rangle_\text{imp} \otimes  e^{-i H_0 t} |\psi_0\rangle,
\end{multline} 
so that the wave function of the system is now entangled with the impurity spin. From Eq.~(\ref{Eq:psi-in-t}) we see that the component of the wave function which has the impurity spin pointing up was evolving with perturbed Hamiltonian, while the part with impurity spin pointing down is evolving with $H_0$. Calculating the expectation value of the impurity spin~$\sigma^x_\text{imp}$ after time $t$,
\begin{equation}\label{Eq:sigma-x}
\langle\Psi(t) |\sigma^x_\text{imp}  |\Psi(t) \rangle
=
\mathop{\rm Re}
 \langle \psi_0 | e^{i(H_0+V) t} e^{-i H_0 t} |\psi_0\rangle,
\end{equation}
we see that it coincides the real part of the overlap $S(t)$ introduced in Eq.~(\ref{Eq:S-def}). 

Hence,  measuring overlap function requires the ability to prepare the system with coupled impurity in a product state, and to observe the expectation value of impurity spin after some time. Both of these requirements are achievable with modern experimental techniques,  motivating the theoretical study of the behavior of overlap function, $S(t)$. While the above considerations were completely general, in what follows we restrict studies of the overlap function to a specific system used as a model of the many-body localized phase. 

More specifically, below we consider XXZ spin chain in a random magnetic field which is defined by the Hamiltonian
\begin{equation}\label{Eq:XXZ}
H_\text{XXZ}=\frac12\sum_{i=1}^{L-1}\left[ \sigma_i^x \sigma_{i+1}^x+\sigma_i^y \sigma_{i+1}^y +J_z\sigma_i^z \sigma_{i+1}^z\right]+\sum_{i=1}^L w_i \sigma_i^z. 
\end{equation}
This model can be in the ergodic or MBL phase, depending on the value of the interaction strength~$J_z$ and disorder strength $W$, which controls the width of the uniform distribution of random fields, $w_i\in [-W,W]$~\cite{OganesyanHuse,PalHuse}. For $J_z=1$ the model is in the many-body localized phase for $W_c\geq 3.7$ even at infinite temperature, i.~e.~all many-body states, even in the middle of the band, are MBL. For weaker values of interaction the critical value of disorder is smaller~\cite{Vasseur14}. Finally, when $J_z=0$ the model~(\ref{Eq:XXZ}) maps onto an Anderson insulator of free fermions and is always localized. 

To fully specify the overlap function, we also need an explicit form of the operator $V$. In what follows we consider the perturbation operator 
\begin{equation}\label{Eq:V-op}
  V =  2g\sigma^z_1,
\end{equation}
where $g$ controls the coupling strength, and $\sigma^z_1$ corresponds to the first spin in the chain. Taking initial Hamiltonian to be $H_0 = H_\text{XXZ}-g\sigma^z_1$, Eq.~(\ref{Eq:S-def}) reduces to a more symmetric form, 
 \begin{equation}\label{Eq:S-symm}
S(t)= \langle \psi_0 | e^{i(H_\text{XXZ}+g\sigma^z_1) t} e^{-i (H_\text{XXZ}-g\sigma^z_1) t} |\psi_0\rangle,
\end{equation}
that will be used below.

In the MBL phase one can diagonalize Hamiltonian~(\ref{Eq:XXZ}) by applying a sequence of quasi-local unitary transformations~\cite{Serbyn13-1,Imbrie16}. The same sequence of quasi-local unitary operators that diagonalizes Hamiltonian can be used to rotate the physical spin operators into local integrals of motion (LIOM) which commute with the Hamiltonian and have exponentially localized support~\cite{Serbyn13-1,Huse13,Imbrie16,Ros14}. Expressed via LIOMs, the Hamiltonian reads:
\begin{equation}\label{Eq:Htau}
H_\text{XXZ}=\sum_{i} h_i \tau^z_i+\sum_{ij} J_{ij} \tau^z_i\tau^z_j+ \sum_{ijk}J_{ijk}\tau^z_i\tau^z_j\tau^z_k +\ldots, 
\end{equation}
where all couplings are exponentially suppressed with the distance $J_{i_1,i_2\ldots,i_k}\propto \exp(-|i_1-i_k|/\xi')$, where we assumed $i_1>i_2>\ldots i_k$.

The physical spin operator also can be expanded over the complete basis of $\tau^\alpha_i$ operators. In particular, we will be interested in the expansion of the perturbation operator~(\ref{Eq:V-op}), given by $\sigma^z_1$. In the basis of LIOM it can be written as 
\begin{multline}\label{Eq:sigma-tau}
\sigma^z_1
=
f^{(0)}[\{\tau^z_i\}]+ f^{(2)}_{kl}[\{\tau^z_i\}]\left(\tau^+_{k}\tau^-_{l}+\text{h.c.}\right)
\\
+ f^{(4)}_{klmn}[\{\tau^z_i\}]\left(\tau^+_{k}\tau^+_{l}\tau^-_{m}\tau^-_{n}+\ldots +\text{h.c.}\right)+\ldots,
\end{multline}
where functions  $f^{(p)}_{i_1,\ldots, i_p}[\{\tau^z_i\}]$ with $p=0,2,\ldots,L$ denote polynomials in $\tau^z$ that couple to terms flipping~$p$ effective spins. For example, 
\begin{equation}\label{Eq:poly-exp}
f^{(0)}[\{\tau^z_i\}]
=
\sum_i c_i \tau^z_i +
\sum_{ij} c_{ij} \tau^z_i\tau^z_j +\ldots,
\end{equation}
where  similarly to the couplings $J_{i_1,i_2\ldots,i_k}$, the coefficients $c_{ij,\ldots}$ decay exponentially with the distance from the site $i_1=1$ where the physical spin is located,
\begin{equation}\label{Eq:c-decay}
 c_i\propto e^{-|i-i_1|/\xi}, 
 \quad
 c_{ij}\propto e^{-\max(|i-i_1|,|j-i_1|)/\xi}, 
 \ldots.
\end{equation}

Recalling that eigenstates correspond to non-entangled configurations $|\uparrow\downarrow\ldots \downarrow\rangle$ of LIOMs, where each spin points either up or down, we may interpret the first term in Eq.~(\ref{Eq:sigma-tau}), $f^{(0)}[\{\tau^z_i\}]$, as being fixed by the \emph{diagonal matrix elements} of operator $\sigma^z_1$ in the basis of eigenstates. All other terms in Eq.~(\ref{Eq:sigma-tau}) label off-diagonal matrix elements which flip progressively larger number of effective spins. Note that the structure of expansion~(\ref{Eq:sigma-tau}) becomes qualitatively different for the non-interacting Anderson insulator. There, the expansion is limited to the terms that contain either $\tau^z_i$ or $\tau_i^+\tau_j^-$ operators. All terms that have more that one $\tau^z$ operator, or flip more than a single spin arise from the interactions. 

\section{Decay of spin coherence with time \label{Sec:overlap}}
In order to understand the behavior of the overlap, it is convenient to transform  Eq.~(\ref{Eq:S-symm}) to the basis of LIOMS using Eqs.~(\ref{Eq:Htau}) and (\ref{Eq:sigma-tau}). Under the assumption that  $\sigma^z_1$ commutes with the Hamiltonian [this is equivalent to retaining only first term in the expansion of $\sigma_1^z$ over LIOMS,  Eq.~(\ref{Eq:sigma-tau})], the overlap becomes:
\begin{equation}\label{Eq:S-commute}
S(t)\approx \langle \psi_0 | e^{2igt f^{(0)}[\{\tau^z_i\}]} |\psi_0\rangle.
\end{equation}
We will discuss and motivate the legitimacy and limitations of such approximation in the next section. 
Assuming weakly entangled initial state, we may approximate the initial state of the spin chain in the LIOM basis as 
\begin{equation}\label{Eq:psi0-tau}
 |\psi_0\rangle
 =
 \otimes_{i =1}^L \left(A_{i\uparrow} |\uparrow\rangle_i +A_{i\downarrow}|\downarrow\rangle_i\right),
\end{equation}
where coefficients $A_{i\sigma}$ depend on the details of the initial state. 

Using explicit form of the initial state and approximated form of $S(t)$ in Eq.~(\ref{Eq:S-commute}), we deduce that the overlap is expressed as a sum of oscillating terms, 
\begin{equation}\label{Eq:S-oscillate}
S(t)
=
\sum_{\{\tau^z\}} \prod_{i=1}^L |A_{i\tau^z_i}|^2 e^{2 i g t \left[\sum_i c_i \tau^z_i +
\sum_{ij} c_{ij} \tau^z_i\tau^z_j+\ldots\right]},
\end{equation}
where the sum runs over all possible $2^L$ configurations of $\{\tau^z\}$ that label entire spectrum of the system. Due to the exponential suppression of couplings $c_{i j\ldots}$ with the range of indices as in Eq.~(\ref{Eq:c-decay}), the dynamics of $S(t)$ will be governed by the slow dephasing mechanism described in Ref.~\onlinecite{Serbyn14}. 

In particular, for time such that $2 g t \leq 1$, the only relevant coupling is $c_1\propto O(1)$, and there are only 2 oscillating terms in the $S(t)$ corresponding to $\tau^z_1=\pm1$. At longer times such that $2gtc_2 \sim 1$, where $c_{2} \propto e^{-2/\xi}$ the second spin begins to matter, and the sum has 4 oscillating terms. Hence, we see that while $S(t)$ will have many oscillating contributions, the number of spins that participate in dephasing grows logarithmically with time. From Eq.~(\ref{Eq:c-decay}) we get that the number of ``dephased'' spins grows as $x = \xi \log 2gt$ (we note that this relation holds when the perturbed spin is at the boundary; if the impurity spin couples to the bulk of the system, there is an extra  factor of 2), so that the number of oscillating terms, $2^x$, will grow as a power-law in time. Collecting all factors, we deduce that fluctuations of $S(t)$ would decay as 
\begin{equation}\label{Eq:S-decay}
  \langle|S(t)|\rangle
  \propto 
  \frac{1}{(2gt)^b}, 
  \quad
  b =\frac12\xi s_2,
\end{equation}
where the power $b$ is related to the second diagonal Renyi entropy density $s_2 = S_2(\ell)/\ell$, and factor of $1/2$ is absent when the perturbed spin is located in the bulk of the system.~\cite{Serbyn14} 
 
Note, that in order to access the dephasing dynamics, it is important to consider the fluctuations of $S(t)$, e.g. by taking the average of the absolute value as in Eq.~(\ref{Eq:S-decay}). If one considers the average overlap $\corr{S(t)}$ without taking the absolute value, as was done in Ref.~\onlinecite{Huber16}, one accesses the generating function of the distribution of $c_i$, instead of the dephasing mechanism, as we show in the Appendix~\ref{Sec:app}.

\begin{figure}[t]
\begin{center}
\includegraphics[width=0.95\columnwidth]{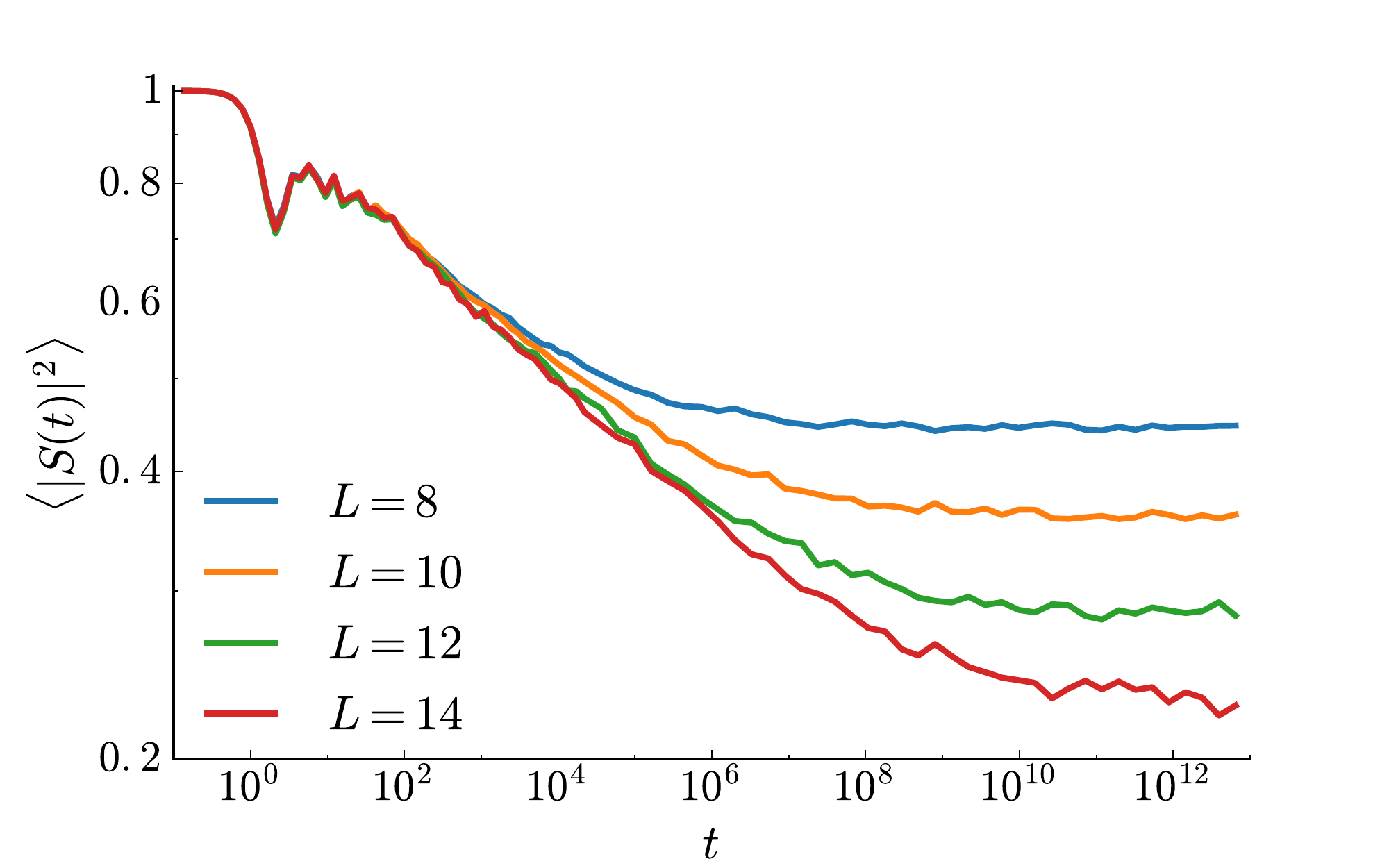}
\caption{ \label{Fig:Ldep} 
Fluctuations of the overlap decay as a power-law in time, saturating to the value that is exponentially suppressed with the system size. The averaging was performed for at least $10^3$ disorder realizations in an XXZ spin chain with disorder $W=6.5$ and interaction $J_z=1$.}
\end{center}
\end{figure}

To illustrate the power-law decay derived above, we calculate the overlap function numerically using exact diagonalization for spin chains up to $L=14$ spins. We start with the spin-density wave state, $|\psi_0\rangle = |\uparrow\downarrow\uparrow\downarrow\ldots\uparrow\downarrow\rangle$, where every even (odd) spin points up (down). Figure~\ref{Fig:Ldep} illustrates the power-law decay of the averaged absolute value of the overlap for the different system sizes. Note, that the saturation value is fairly large even for the system of $L=14$ spins, which is naturally explained by the strong value of disorder $W=6.5$ and  initialization of the system in the Neel state at $t=0$. 

\begin{figure}[t]
\begin{center}
\includegraphics[width=0.95\columnwidth]{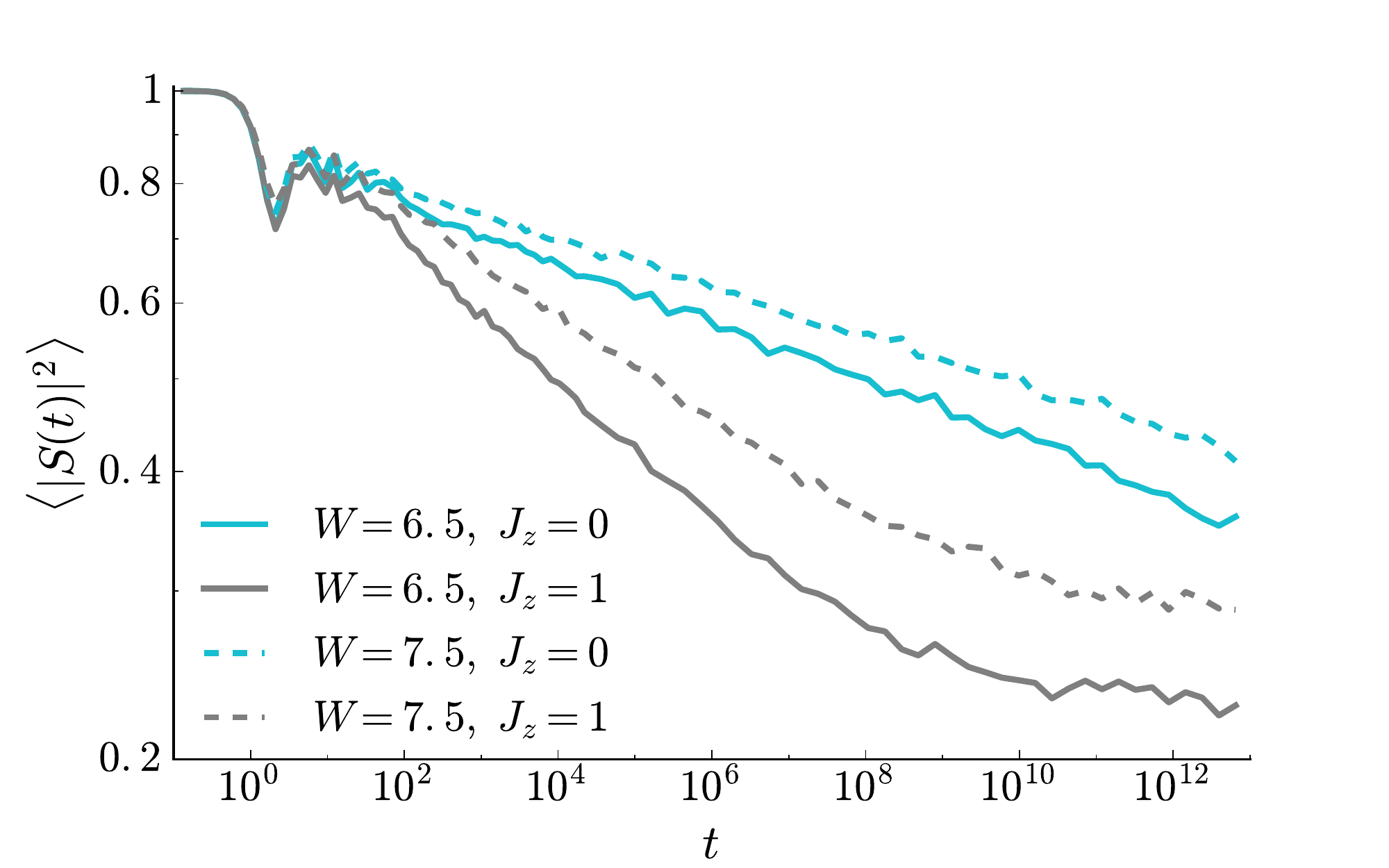}\\
\caption{ Cyan and grey lines show the power law decay of $\corr{|S(t)|}$ in Anderson insulator and MBL phase. Note the much faster overlap decay in the MBL phase. Moreover, the exponent of the decay is more sensitive to the increase in the value of disorder in the MBL phase. System has~$L=14$ spins.
\label{Fig:WJdep} 
}
\end{center}
\end{figure}

It is instructive to compare the decay of the overlap in the MBL phase to the case of Anderson insulator. Figure~\ref{Fig:WJdep} illustrates that the decay becomes slower with increased value of disorder. This is indeed what Eq.~(\ref{Eq:S-decay}) predicts, because for stronger disorder the Neel state has progressively larger overlap with an exact eigenstate, hence diagonal Renyi entropy density $s_2$ goes down, leading to slower decay of $S(t)$. Decreasing interaction strength has similar effect, but affects the decay even stronger. Note that in the non-interacting case there are only linear in $\tau^z$ terms in the exponent in Eq.~(\ref{Eq:S-oscillate}), while non-zero value $J_z$ leads to the presence of operators with support on many spins. Hence, while non-zero interaction weakly impacts the saturation value of imbalance, it is the change in the structure of the operator expansion that is causing faster overlap decay in the presence of interactions.

\section{Impurity spin echo protocols \label{Sec:echo}}

Next, let us return to the approximation made in the beginning of the previous section, where we neglected terms that flip effective spins in the expansion of the $\sigma^z_1$ operator. Such terms can be conveniently probed in a spin echo type protocol performed on the impurity. Namely, if one applies a $\pi$-pulse to the impurity at time $t$, and allows the system to evolve for an additional time $t$ before measuring $\sigma^x_\text{imp}$, this gives access to the real part of the following expectation value:
\begin{subequations}\label{Eq:S-echo}
\begin{eqnarray}
  S_\text{echo}(t)
  &=&
  \corr{\psi_0|U_\text{echo}(t)|\psi_0},\\ \label{Eq:U-echo}
 U_\text{echo}(t) &=& 
 e^{i(H_0+V)t} e^{i H_0t}e^{-i(H_0+V)t} e^{-iH_0t}
\end{eqnarray}
\end{subequations}
The overlap, defined in  Eq.~(\ref{Eq:S-def}) was measuring the similarity between the wave function evolved with perturbed and unperturbed Hamiltonian. In contrast, the spin-echo protocol, Eq.~(\ref{Eq:S-echo}) probes the overlap between wave functions which are evolved with both, perturbed and unperturbed Hamiltonian, but the \emph{order of the evolution} is reversed between the two. 

In order to understand the behavior of $S_\text{echo}(t)$, it is convenient to rewrite the unitary operator in Eq.~(\ref{Eq:U-echo}) as follows:
\begin{equation}\label{Eq:U-rewrite}
  U_\text{echo}(t)=  e^{i(H_0+V)t} e^{-i(H_0+V[t]_0)t},
\end{equation}
where we promoted operators $e^{\pm iH_0t}$ inside the exponent, and introduced  short-hand notation $V[t]_0$ for operator $V$ time-evolved with Hamiltonian $H_0$:
\begin{equation}\label{Eq:V-time}
V[t]_0=e^{i H_0t} V e^{-iH_0t}.
\end{equation}

\begin{figure}[t]
\begin{center}
\includegraphics[width=0.95\columnwidth]{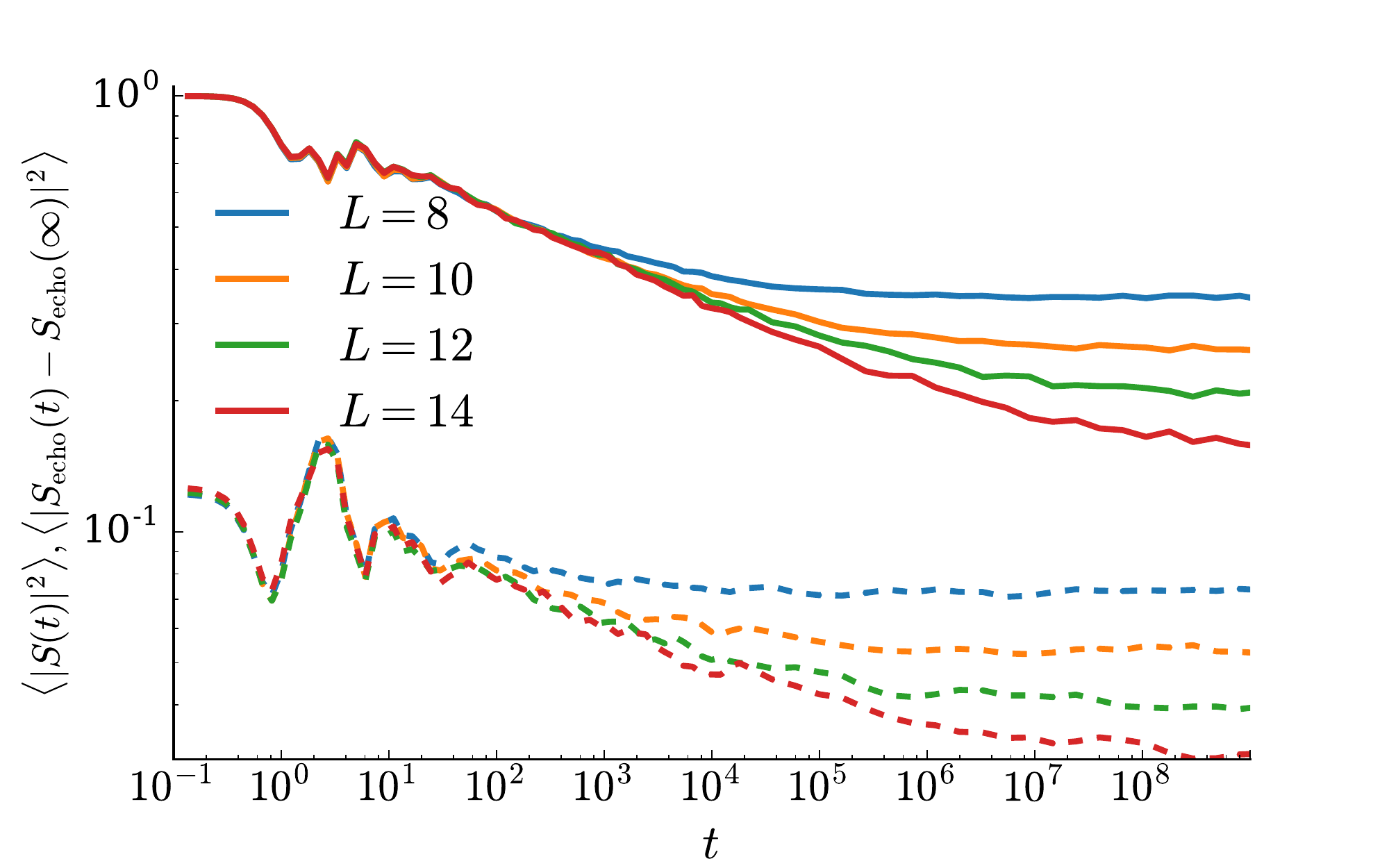}\\
\caption{ Fluctuations of impurity spin echo signal decay in a similar way to the overlap (solid lines). Value of disorder is $W=5$, interaction $J_z=1$, $g=2$.
\label{Fig:echo} 
}
\end{center}
\end{figure}

Using Eq.~(\ref{Eq:U-rewrite}), we may reinterpret the $S_\text{echo}(t)$ as an overlap between wave function that is evolved with two different Hamiltonians $H_1=H_0+V$ and $H_2=H_0+V[t]_0$. These two Hamiltonians have identical spectra, since they are related as $H_2=H_1[t]_0$, and spectrum remains invariant under evolution with an arbitrary unitary operator. Due to the identical spectra of $H_1$ and $H_2$, the decay of $S_\text{echo}(t)$ can be caused only by the difference in the eigenbasis of these Hamiltonians, which in turn depends on the difference between $V_0$ and time-evolved $V[t]_0$ operators. 

The difference between operator $V_0$ and its time-evolved version crucially depends on the presence of interactions in the system. If there is no interactions, $J_z=0$, and system is an Anderson insulator, the operator $V[t]_0$ does not spread beyond single-particle localization length, and it always remains localized. Hence, in the non-interacting case we do not expect to see any decay of fluctuations of $S_\text{echo}(t)$ on long time scales. 

On the other hand, in the many-body localized phase the operators spread logarithmically in time. From the expansion~(\ref{Eq:sigma-tau}) it is straightforward to work out the time-evolved form of the operator $\sigma^z_1$. Under evolution with Hamiltonian~(\ref{Eq:Htau}) diagonal terms remain invariant, while off-diagonal terms acquire arbitrary long ``tails'' consisting of $\tau^z$ operators. For instance, the time-evolved operator $\tau^x_1$ that is contained within expansion~(\ref{Eq:sigma-tau}) becomes:
\begin{equation}\label{Eq:taux-time}
 \tau^x_1[t]_0
 =
 \cos (2 H_1[\{\tau^z_i\}]t ) \tau_1^x-
  \sin (2 H_1[\{\tau^z_i\}]t ) \tau_1^y,
\end{equation}
where $H_1[\{\tau^z_i\}]$ is an (operator) magnetic field experienced by the first spin that is given by the linear in $\tau^z_1$ term in Eq.~(\ref{Eq:Htau}). This magnetic field contains a local field, two-spin terms, and so on,
\begin{equation}\label{Eq:H1}
H_1[\{\tau^z_i\}]=
h_1+\sum_i {}' J_{1i}\tau^z_i+\sum_{ij} {}'J_{1ij}\tau^z_i\tau^z_j+\ldots,
\end{equation}
where prime denotes that indices are not repeated and are different from $1$, $i,j\neq 1$, see Ref.~\onlinecite{Serbyn14} for more details. 

Due to exponential hierarchy of couplings $J_{ij,\ldots}$ the number of terms that are relevant  in Eq.~(\ref{Eq:H1}) grows logarithmically with time, causing a logarithmic growth of~$ \tau^x_1[t]_0$. For example, leaving only nearest and next-nearest-neighbor two-spin terms we get $H_1[\{\tau^z_i\}]\approx h_1+J_{12}\tau^z_2+J_{13}\tau^z_3$, leading to  
\begin{multline}\label{Eq:cos-ex}
 \cos (2 H_1[\{\tau^z_i\}]t ) 
 =
 {\rm c}_{1}  {\rm c}_{12}  {\rm c}_{13}-  {\rm s}_1 {\rm s}_{12}  {\rm c}_{13}\, \tau^z_2
\\
 -  {\rm s}_1 {\rm c}_{12}  {\rm s}_{13}\, \tau^z_3
  -  {\rm c}_1 {\rm s}_{12}  {\rm s}_{13}\, \tau^z_2\tau^z_3,
\end{multline}
where we introduced  short-hand notations ${\rm c}_{i} = \cos (2h_{i}t)$, ${\rm c}_{ij} = \cos (2J_{ij}t)$ and  ${\rm s}_{i} = \sin (2h_{i}t)$, ${\rm s}_{ij} = \sin (2J_{ij}t)$. From here we see that at times such that $J_{12}t \sim 1$, the operator $ \tau^x_1[t]_0$ acquires a term $ \tau^x_1 \tau^z_2$, while at longer times when  $J_{13}t \sim 1$ two more terms  emerge, including $ \tau^x_1 \tau^z_2\tau^z_3$.  Eventually at sufficiently long times the $\tau^x$ operator will include terms  
\begin{equation}\label{Eq:taux-evolve}
\tau^{x,y}_1 \tau^z_2,
\quad 
 \tau^{x,y}_1 \tau^z_2\tau^z_3,
 \quad
 \ldots,\quad 
 \tau^{x,y}_1 \tau^z_2\tau^z_3\cdots \tau^z_L.
\end{equation}

From above example we observed that all spin-flip terms in the expansion of $\sigma^z_1$ develop long strings of $\tau^z$ operators with time. Nonetheless, these $\tau^z$ strings cannot flip any LIOM spins. Hence, while operator $\sigma^z_1[t]_0$ spreads up to the full system size, spin flip terms \emph{remain localized} in vicinity of site $i=1$. Physically, this can be interpreted as a fact that local operator can produce excitations only within a finite region, but energy of those excitations in the MBL phase depends on the state of all other spins in the system. Thus, such operator spreading is qualitatively different from the one in the ergodic phase. There, the time-evolved local operator is able to produce excitations throughout the entire volume of the system. 

After understanding the operator spreading, we can return to the discussion of spin-echo overlap. As we demonstrated, the operator spreading causes the eigenbases of $H_1$ and $H_2$ to be different from each other. Nevertheless, due to the fact that time-evolved operator in the MBL phase still can produce only local excitations, we expect the finite saturation value of the spin-echo overlap. In the Appendix~\ref{Sec:app2} we calculate the saturation value of spin-echo by expanding the expression for the spin-echo signal, Eq.~(\ref{Eq:S-echo}) in the eigenstate basis of $H_0+V$. This value is given by the second participation ratio of the eigenstates of $H_0$, denoted as $| \lambda_i\rangle$ over eigenstates of perturbed Hamiltonian, $| \tilde \lambda_j\rangle$:
\begin{equation}\label{Eq:echo-sat}
 S_\text{echo}(\infty) =\frac{1}{\cal D}\sum_{j,i} |\corr{\lambda_i|\tilde \lambda_j}|^2,
\end{equation}
where $\cal D$ is the Hilbert space dimension. While in the ergodic phase such participation ratio would be exponentially suppressed in the system size,  in our many-body localized system this participation ratio is finite~\cite{Serbyn13-1}. Hence, the spin-echo signal relaxes towards a finite value that does not depend on the system size. 

On the other hand, the operator spreading leads to the relaxation of the \emph{fluctuations} of spin-echo overlap. The support of the operator $V[t]_0$ grows as $x_V(t) = \xi' \log J_z t $. While this operator still produces only local excitations, the energy of these excitations depend on state of $x_V(t)$ spins that increases due to accumulation of long $\tau^z$ strings in the dynamics. Assuming the initial state similar to Eq.~(\ref{Eq:psi0-tau}), we obtain the same dephasing mechanism, that now relaxes the fluctuations of spin-echo. The number of oscillating terms grows exponentially with $x_V(t)$, and we expect the fluctuations of the spin echo around its saturation value to decay as
\begin{equation}\label{Eq:S-echo-decay}
  \langle |S_\text{echo}(t)-S_\text{echo}(\infty)|^2 \rangle
  \propto 
  \frac{1}{t^{b'}}, 
  \quad
  b' =\frac12\xi' s_2.
\end{equation}
Note that exponent is again controlled by the second diagonal Renyi entropy density, and the scale~$\xi'$ that controls entanglement spreading and operator growth. At the same time, we would like to point out that the decay of the fluctuations does not imply  transport of conserved quantities (in particular, spin density that is conserved in XXZ spin chain), as spin flip terms remain localized in $V[t]_0$.

To illustrate the results, we present numerical studies of fluctuations of $S_\text{echo}(t)$ in Fig.~\ref{Fig:echo}, comparing it with the usual overlap within the MBL phase. We again use the symmetrized form, measuring fluctuations of the following quantity: 
 \begin{multline}\label{Eq:S-echo-spin}
S_\text{echo}(t)= \langle \psi_0 | e^{i(H_\text{XXZ}+g\sigma^z_1) t} e^{i (H_\text{XXZ}-g\sigma^z_1) t} \\
\times e^{-i(H_\text{XXZ}+g\sigma^z_1) t} e^{-i (H_\text{XXZ}-g\sigma^z_1) t} |\psi_0\rangle. 
\end{multline}
We note, that the initial decay of the spin-echo overlap itself (not shown) is faster compared to the decay of $S(t)$. The rapid decay is caused by the dynamics on the length scales below the localization length, and it fully agrees with the intuition provided in Ref.~\onlinecite{Knap12} that the spin-echo exponent is larger compared to the exponent for the usual overlap in the system of free fermions without disorder. 

\begin{figure}[t]
\begin{center}
\includegraphics[width=0.95\columnwidth]{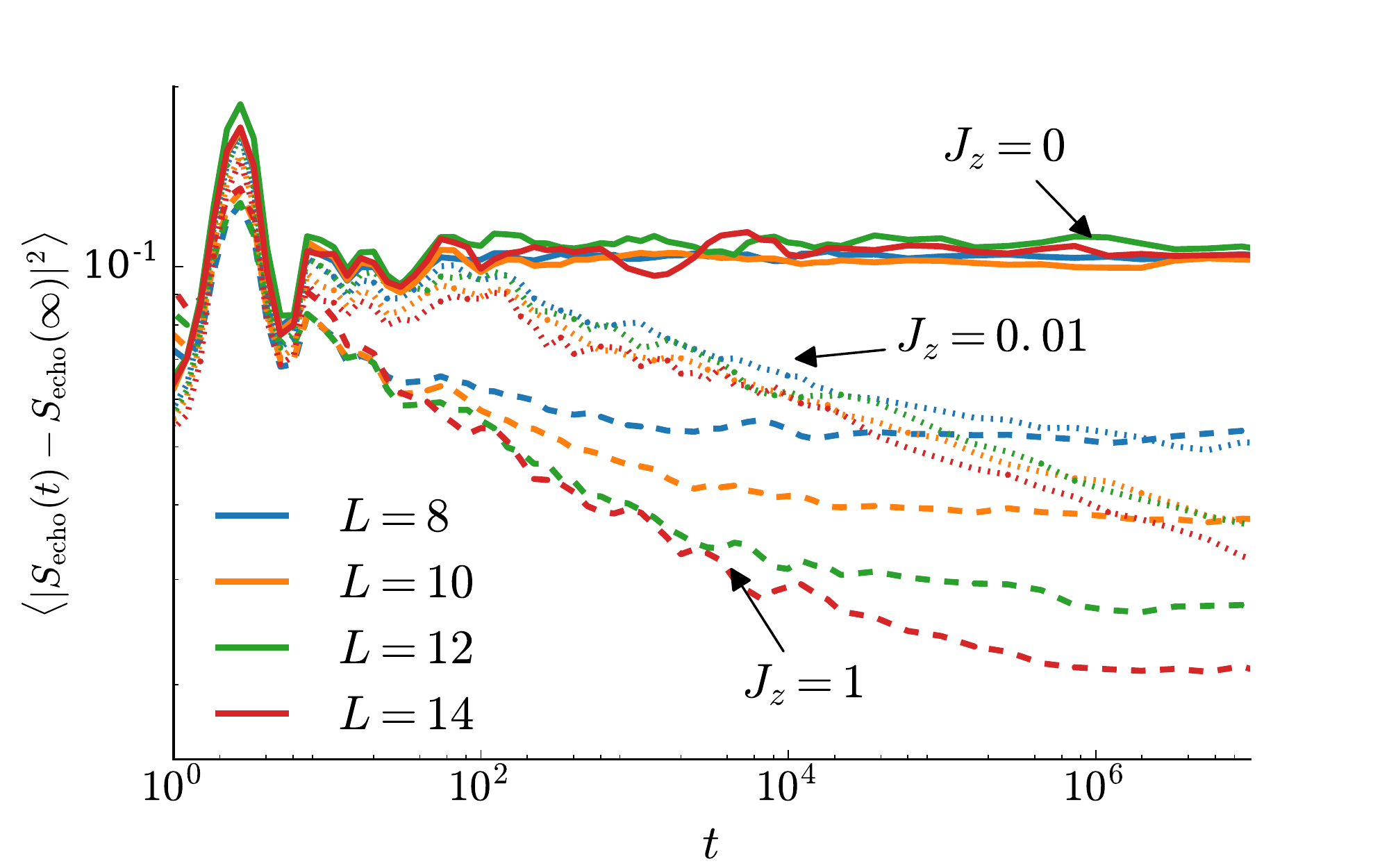}\\
\caption{ \label{Fig:echo-Jz}
Fluctuations of the spin-echo overlap do not relax in the non-interacting system at long-times (solid curves), while presence of even small interactions ($J_z=0.01$, dotted lines) leads to a slow power-law like decay and a residual fluctuations that decrease exponentially with the system size.  Increasing interaction strength to $J_z=1$ (solid lines) gives even faster decay of spin-echo overlap.  Data is obtained for disorder $W=4$ and perturbation strength is $g=4$.}
\end{center}
\end{figure}

On longer length scales our system is localized, and different physics comes into play. The spin echo \emph{saturates} to the finite value (not shown), while its fluctuations slowly relax, see Fig.~\ref{Fig:echo}. Note that  the decay of fluctuations of spin echo is very similar to the decay of the fluctuations of overlap, suggesting that $\xi \approx \xi'$ in Eqs.~(\ref{Eq:S-decay}) and (\ref{Eq:S-echo-decay}). Finally, Fig.~\ref{Fig:echo-Jz}  illustrates the dependence of the spin-echo fluctuations  decay on the interaction strength. In particular, fluctuations do not relax when $J_z=0$. When $J_z\neq 0$, the saturation value of the fluctuations has a weak dependence on the interaction strength, similarly to the fluctuations of overlap $S(t)$.

\section{Summary and outlook \label{Sec:summary}}

In conclusion, we studied the behavior of the Loschmidt echo and its spin-echo generalization in the localized phase with and without interactions. We demonstrated that the  \emph{fluctuations} of the overlap function have a power-law decay both in Anderson insulator and MBL phase. The power-law decay can be contrasted with the exponential decay of the Loschmidt echo in  ergodic systems, reflecting extreme sensitivity of the unitary dynamics of the ergodic systems to the local perturbation. This can be viewed as yet another signature of the non-ergodic dynamics in the MBL phase.

Let us discuss the differences between the overlap, which is decaying irrespectively of the presence of interactions,  
and, for example, fluctuations of the local observables, which do not relax in the Anderson insulator, while decaying as a power-law in the MBL phase~\cite{Serbyn14}. The power-law decay of the orthogonality catastrophe overlap in the Anderson insulator closely parallels the dynamics of the entanglement propagation. The earlier probes, such as relaxation of local observables~\cite{Serbyn14} or quantum revivals~\cite{Vasseur14}  considered the unitary evolution with a fixed Hamiltonian. Such evolution does not cause entanglement spreading in the non-interacting systems,~\cite{Moore12,Serbyn13-2} hence explaining perfect revivals and absence of relaxation of local observables. In contrast, the orthogonality catastrophe setup can be interpreted as the sequential forward and backward evolution of initial state $|\psi_0\rangle$ with two \emph{different} Hamiltonians, $H_0$ and $H_0+V$, which generically have different spectra (single-particle energies). The difference in spectra between $H_0$ and $H_0+V$ gives rise to dephasing mechanism and entanglement growth even in the absence of interactions. 

While the decay of the overlap is qualitatively similar in the Anderson insulator and MBL phase, the exponent of the decay is  sensitive to the presence of interactions. Hence, the orthogonality catastrophe setup can be used to probe the decay of the diagonal (e.g. commuting with the Hamiltonian) part in the expansion of the perturbation operator $V$ over LIOMs. 

Next, we would like to highlight the differences between the physics probed by the overlap function~(\ref{Eq:S-def}) with the conventional orthogonality catastrophe physics. In the original work by Anderson, the orthogonality catastrophe was defined as the effect of the single impurity on the ground state of the Fermi gas~\cite{Anderson67}. These results imply that in the absence of disorder for the initial state $|\psi_0\rangle$ being a filled Fermi sea,  the overlap $|S(t)|^2$ decays as a power-law in time with an exponent set by the scattering phase of impurity potential~\cite{Anderson67}. In contrast, in the present work we consider disordered systems, where all eigenstates are localized, and decay occurs via dephasing. In particular, if we  initialize our system in an eigenstate of $H_0$ or $H_0+V$, the fluctuations would not decay. For the dephasing mechanism to be at play it is important to start with the initial state $|\psi_0\rangle$ that is a superposition of many eigenstates. 

In addition to the standard  orthogonality catastrophe, we also considered the spin-echo type overlap function. In particular, we demonstrated that it singles out and allows to probe the \emph{off-diagonal} (spin-flip) terms in the operator expansion of $V$ over LIOMs. In the Anderson insulator the spin-echo overlap has no dynamics: diagonal terms do not contribute to the spin-echo setup, while off-diagonal terms remain local. The presence of arbitrary small interactions qualitatively changes the dynamics of the spin-echo overlap. Now, the logarithmic in time spreading of the off-diagonal parts of $V$ causes the power-law decay of spin-echo overlap fluctuations. In this sense it is interesting to draw the parallel between spin-echo type overlap and out-of-time ordered correlation function recently demonstrated to have a power-law decay in the MBL phase~\cite{Huang16,*Fan16,*Chen16,*He16,*Swingle16,HuseOTOC}. 

The setup for measuring orthogonality catastrophe and its spin-echo extension works for generic initial non-equilibrium states, and requires only manipulation of the local degrees of freedom. Hence it can be potentially implemented in  systems of cold atoms in optical lattices and trapped ions, where signatures of MBL phase were recently observed. Nevertheless, one has to be able to access the fluctuations of the local observables in order to probe the dephasing dynamics, as the naive averaging of the observables probes different physics (see Appendix~\ref{Sec:app}). Provided one has access to the fluctuations, measurements of  orthogonality catastrophe and spin-echo overlap could be useful for exploring structure of the expansion of local operators over LIOMs in the MBL phase. 

\section*{Acknowledgments}
This research was supported in part by the National Science Foundation under Grant No. NSF PHY11-25915. M.S. was supported by Gordon and Betty Moore Foundation's EPiQS Initiative through Grant GBMF4307. D.A. also acknowledges support by Swiss National Science Foundation.

\appendix
\section{Understanding time-averaged coherence \label{Sec:app}}
In this Appendix we consider the behavior of the overlap decay averaged over disorder realizations, $|\langle S(t)\rangle|$. From Eq.~(\ref{Eq:S-oscillate}) it is clear that the overlap averaged over disorder realizations depends on the distribution of the coefficients $c_{i}, c_{ij},\ldots$. For simplicity, let us ignore the effect of interaction. Then it is legitimate to keep only the leading order coefficients in Eq.~(\ref{Eq:S-oscillate}), and we deduce 
\begin{equation}\label{Eq:St-average}
 \corr{S(t)} =  \corr{\sum_{\{\tau\}}\prod_{i=1}^L (|A_{i\uparrow}|^2 e^{2igt c_i}+|A_{i\downarrow}|^2 e^{-2igt c_i})}.
\end{equation}
From here, neglecting the correlation between $A_{i\tau}$ and $c_i$, we see that the time-dependence of the $\corr{S(t)}$ comes from the disorder-averaged $e^{2igt c_i}$, which is determined by the characteristic function (or, equivalently, Fourier transform) of the distribution of $c_i$, $p(c_i)$,
\begin{equation}\label{Eq:P-gen}
\corr{e^{2igt c_i}} =  \int dc_i\,  p(c_i) e^{2igt c_i} =  \varphi_{c_i}(2gt).
\end{equation}

In the non-interacting case the coefficients $c_i$ in the expansion~(\ref{Eq:sigma-tau}) are given by the tails of the single-particle wave function. Using log-normal form of the distribution of the inverse localization length~\cite{MirlinRMP}, we replace  $ \varphi_{c_i}(t)$ with an asymptotic form of the characteristic function of the log-normal distribution~\cite{Asmussen2016}
\begin{equation}\label{Eq:LN-char-fun}
 \varphi_{c_i} (t)\approx {\frac {\exp \left(-{\frac {W^{2}(t\sigma_i ^{2}\corr{c_i})+2W(t\sigma_i^{2}\corr{c_i})}{2\sigma_i ^{2}}}\right)}{\sqrt {1+W(t\sigma_i ^{2}\corr{c_i})}}},
\end{equation}
where $W$ is the Lambert W-function, $\corr{c_i}$ is the median (typical) value of the corresponding coefficient in the expansion, and $\sigma_i$ is the variance of $\ln c_i$. 

\begin{figure}[t]
\begin{center}
\includegraphics[width=0.95\columnwidth]{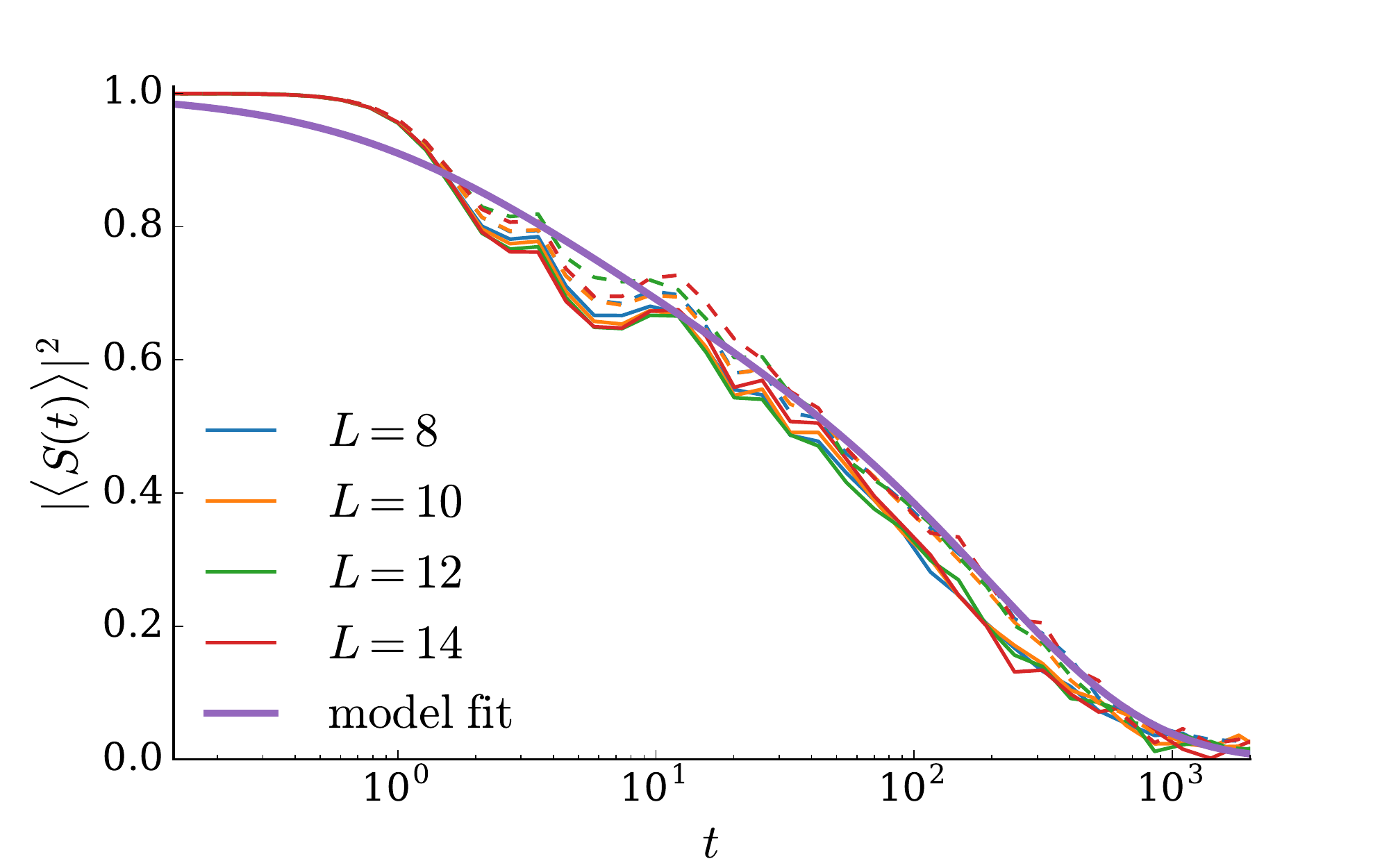}\\
\caption{ \label{Fig:Maverage} 
Absolute value of the averaged coherence does not depend on the system size and interaction strength, and has a weak dependence on the disorder strength (solid lines correspond to $W=6.5$ and dashed lines to $W=7.5$). The numerical data agrees reasonably well with the theory suggesting the log-normal distribution of the localization length.}
\end{center}
\end{figure}

Using expression~(\ref{Eq:LN-char-fun}), we can approximate the average overlap as 
\begin{equation}\label{Eq:St-approx}
\corr{S(t)} \approx  \prod_{i=1}^L  \varphi_{c_i} (2gt).
\end{equation}
Practically, the above product quickly converges since each $c_{i+1}$ is suppressed compared $c_i$ by an extra factor of $e^{-1/\xi}$, and it can be truncated at $i=2$. Hence, the expression~(\ref{Eq:St-approx}) has only three independent parameters: median value of $c_1$, its variance, $\sigma_1$, and suppression factor $e^{-1/\xi}$.

Treating $\corr{c_1}$, $\sigma_1$, and $e^{-1/\xi}$ as fitting parameters, we compare the predictions of Eq.~(\ref{Eq:St-approx}) to the numerical data obtained for the XXZ spin chain in Fig.~\ref{Fig:Maverage}. The numerical data weakly depends on the value of disorder, and shows almost no dependence on the system size and interaction strength, consistent with the convergence of the product in Eq.~(\ref{Eq:St-approx}). Fit with Eq.~(\ref{Eq:St-approx}), shown in Fig.~\ref{Fig:Maverage} adequately reproduces the time dependence of $|\corr{S(t)} |$ at intermediate time. 

Physically, the non-trivial behavior of the $|\corr{S(t)} |$ with time arises from the broad distribution of the coefficients~$c_i$ that determine the distribution of the oscillation frequencies. Quick convergence of the product in Eq.~(\ref{Eq:St-approx})  explains why the interactions do not affect the dependence of $|\corr{S(t)} |$ on the intermediate times: both terms with $c_i$ with $i>1$, and terms involving more $\tau^z$ in Eq.~(\ref{Eq:poly-exp}) are exponentially suppressed. Hence, their effect is not important on the timescales shown in Fig.~\ref{Fig:Maverage}. 

\section{Decay of spin-echo overlap and entanglement dynamics\label{Sec:app2}}

Below we consider the behavior of the spin-echo overlap $S_\text{echo}(t)$. In the main text we argued the decay of this overlap as originating from spreading of operator $V[t]_0$ defined in Eq.~(\ref{Eq:V-time}) with time. However, it is instructive to consider the decay of  $S_\text{echo}(t)$ from the perspective of eigenstate dynamics. For this we expand the initial state over eigenstates of operator $H_0+V$, as 
\begin{equation}\label{Eq:eig-exp}
|\psi_0\rangle
=
\sum_i \alpha_i |\tilde\lambda_i\rangle,
\end{equation}
where the sum involves a number of eigenstates that is proportional to the size of the Hilbert space. Eigenstates $|\tilde\lambda_i\rangle$ are assumed to have energy $\tilde\lambda_i$. Using this representation, we rewrite the overlap using the fact that  eigenstates of $H_0+V$ only acquire a phase under action of $ e^{i(H_0+V)t} $:
\begin{subequations}\label{Eq:S-echo-es}
\begin{eqnarray}\label{Eq:S-echo-es-1}
  S_\text{echo}(t)&=&
  \sum_{i,j} \alpha^*_i \alpha_j s_{ij}(t),\\\label{Eq:S-echo-es-2}
 s_{ij}(t)&=& e^{i\tilde\lambda_it}\corr{\tilde\lambda_i|  e^{i H_0t}e^{-i(H_0+V)t} e^{-iH_0t}|\tilde\lambda_j},
\end{eqnarray}
\end{subequations}
where we defined as spin-echo response of a pair of eigenstates $i$ and $j$, $s_{ij}(t)$. Further, we expand eigenstates of Hamiltonian $H_0+V$ over eigenstates of $H_0$. Since these two Hamiltonians are related by the local perturbation and system is in the many-body localized phase, each eigenstate $|\tilde\lambda_i\rangle$ can be expressed as a sum of \emph{finite} number of eigenstates of unperturbed Hamiltonian, $|\lambda_i\rangle$ (up to exponentially small corrections):
\begin{equation}\label{Eq:eig-exp-2}
|\tilde\lambda_i\rangle
\approx 
\sum_{k \in I_i} u_{ik} |\lambda_k\rangle,
\quad
|\lambda_k\rangle
\approx 
\sum_{i \in \tilde I_k} u^*_{ik} |\tilde\lambda_i\rangle,
\end{equation}
where sets $I_i$, $\tilde I_k$ depend on corresponding eigenstates $|\tilde\lambda_i\rangle$, $|\lambda_k\rangle$, and include a finite number of indices. This follows from the local effect of the local perturbation in the MBL phase; the similar participation ratios were explicitly calculated in Ref.~\onlinecite{Serbyn13-1}. 

Applying the expansion~(\ref{Eq:eig-exp-2}) twice, we get the following result for the $s_{ij}(t)$:
\begin{multline}\label{Eq:corr-2}
s_{ij}(t)=
e^{i\tilde\lambda_it}\hspace{-8pt}\sum_{k\in I_i,n\in I_j}
e^{i(\lambda_k-\lambda_n)t}u^*_{ik}u_{jn}\corr{\lambda_k|  e^{-i(H_0+V)t} |\lambda_n}
\\=
\hspace{-5pt}\sum_{k\in I_i,n\in I_j,q\in \tilde I_k \cap \tilde I_n}
e^{i(\lambda_k-\lambda_n+\tilde\lambda_i-\tilde \lambda_q)t}u^*_{ik}u_{qk}u^*_{qn} u_{jn} .
\end{multline}
If operator $V$ had no off-diagonal matrix elements, the unitary matrix $u_{ik}$ would be the permutation matrix, having only single non-zero element in each row/column.  In this case the $S_\text{echo}(t)$ would always remain equal to one. Presence of off-diagonal matrix elements in operator $V$ leads to the decay of the expectation value~(\ref{Eq:corr-2}).  

Nevertheless, the expectation value of diagonal operators $s_{ij}(t)$ with $i=j$ saturates to a finite value $\bar s_{ii}= s_{ii}(t\to\infty)$ that does not scale with the system size. This saturation value is given by the terms in the sum in Eq.~(\ref{Eq:corr-2}) that have no oscillations in time, which corresponds to the part with $k=n$ and $i=q$: 
\begin{equation}\label{Eq:corr-2-sat}
\bar s_{ii}=
e^{i\tilde\lambda_it}\sum_{k\in I_i}
u^*_{ik}u_{ik} u_{ik}u^*_{ik} e^{-i\tilde \lambda_i t}
=
\sum_{k\in I_i}
|u_{ik}|^4,
\end{equation}
where all oscillating terms cancel. From here we see that $\bar s_{ii}$ is given by a second participation ratio of the eigenstates of $H_0$ in the basis of perturbed Hamiltonian $H_0+V$. Finite value of $\bar s_{ii}<1$ translates into the finite saturation value of $S_\text{echo}(t)$ at long times, as $\sum_i|\alpha_i|^2=1$ in Eq.~(\ref{Eq:S-echo-es-1}). Note, that this result implies a weak dependence of the saturation value of spin-echo on the choice of initial state.

\begin{figure}[t]
\begin{center}
\includegraphics[width=0.95\columnwidth]{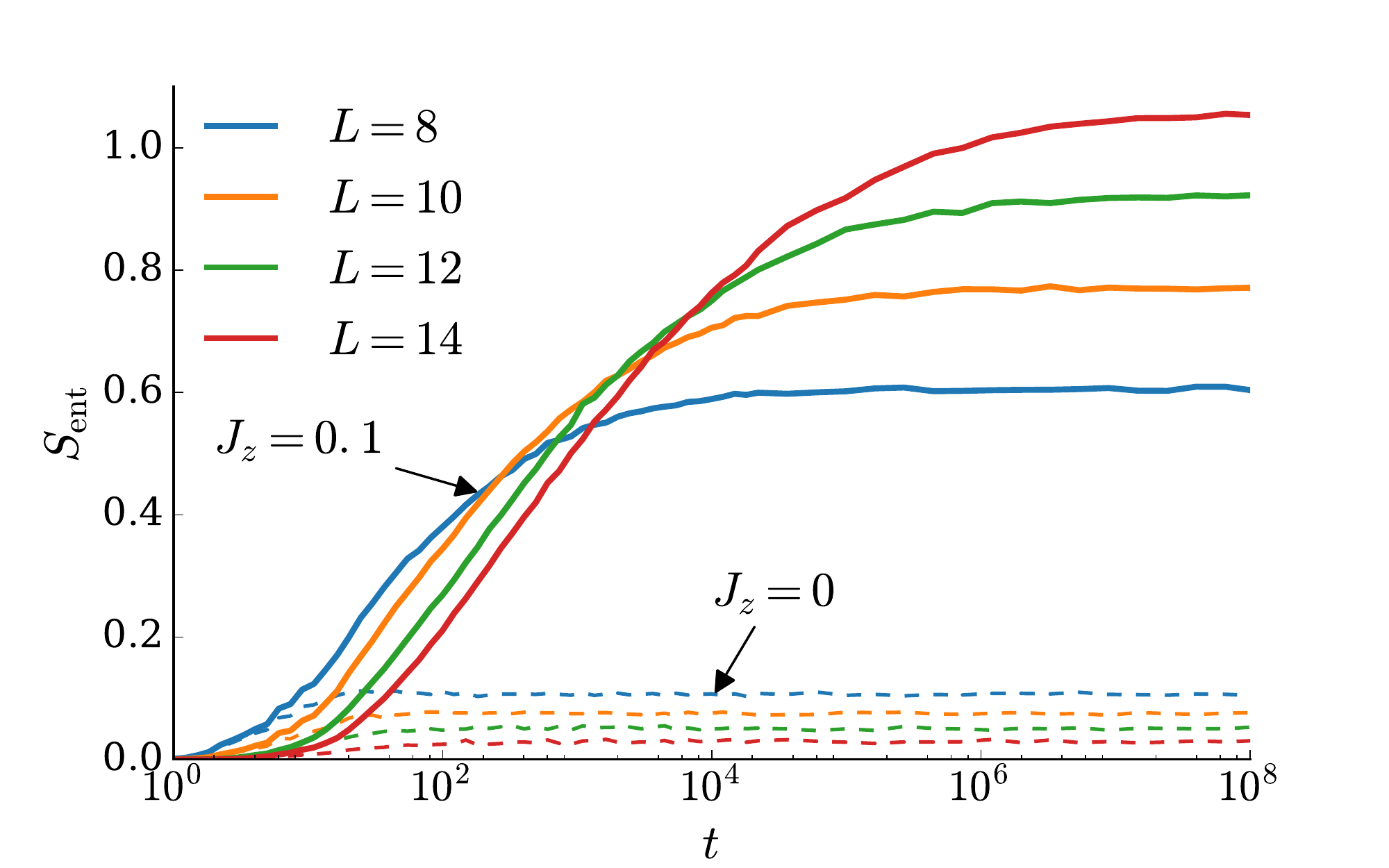}\\
\caption{ \label{Fig:ent}
Entanglement generated by $U_\text{echo}(t)$ across the middle link of the spin chain depends on the presence of interactions. When $J_z>0$ entanglement grows logarithmically and has extensive saturation value (solid lines). In contrast, in the non-interacting system saturation value of entanglement decreases with the system size (dashed lines). Disorder strength is $W=3$, and $g=4$.
}
\end{center}
\end{figure}

Above we demonstrated that diagonal terms in Eq.~(\ref{Eq:S-echo-es}) are responsible for the finite saturation value of spin-echo. At the same time these terms do not contribute to the relaxation of spin-echo fluctuations. Due to local character of operator $V$ the range of summation in Eq.~(\ref{Eq:corr-2}) is restricted~(sets $I_{i}$ and others include a number of indices that does not depend on the system size), hence fluctuations of individual $s_{ii}(t)$ do not relax. 

On the other hand, the spin echo overlap Eq.~(\ref{Eq:S-echo-es}) generally contains an extensive number of off-diagonal terms $s_{ij}(t)$ with $i\neq j$. Hence, even though operator $V$ is able to relate each eigenstate only to a finite number of other eigenstates by producing local excitations, the fluctuations of \emph{different} $s_{ij}(t)$ relax via dephasing mechanism~\cite{Serbyn14,Serbyn_14_Deer}. More specifically, the $s_{ij}(t)$ can be non-zero only if eigenstates $\tilde \lambda_i$ and $\tilde \lambda_j$ are different in vicinity of operator $V$.  However, the energy difference in the exponent in Eq.~(\ref{Eq:corr-2}) depends on the state of all spins in the system. In other words, the energies of same local excitation for different eigenstates would be split by an exponentially small amount depending on the state of the distant spins~\cite{Nandkishore14}. This splitting, described in the main text via the operator spreading gives rise to oscillations at a sufficiently long times, and leads to the power-law decay of spin-echo fluctuations.

Finally, we illustrate the entanglement dynamics under the action of the unitary operator $U_\text{echo}(t)$, defined in Eqs.~(\ref{Eq:U-echo}). Taking $|\psi_0\rangle$ to be the Neel state, we have no entanglement at $t=0$. Figure~\ref{Fig:ent} shows the entanglement entropy of state $U_\text{echo}(t)|\psi_0\rangle$ as a function of time~$t$. The entanglement cut is at the middle link of the system. Note, that in the non-interacting case there is no entanglement growth at long times. Moreover, the saturation value of entanglement at the middle link decreases with the system size, as the distance between the entanglement cut and site where perturbation $V$ is applied increases with system size as $L/2$. This confirms that the operator $V[t]_0$ remains local in the Anderson insulator, and goes in parallel with the absence of the decay of fluctuations of  $S_\text{echo}(t)$. In contrast, presence of weak interactions qualitatively changes the entanglement dynamics, which now spreads logarithmically in time. The saturation value of entanglement is proportional to the system size. At the same time,  fluctuations of $S_\text{echo}(t)$ now also have a non-trivial decay, emphasizing that $\corr{|S_\text{echo}(t)|^2}$ probes the same physics.

\end{document}